\documentclass[a4paper,11pt]{article}
\pdfoutput=1
\usepackage{amsmath}
\usepackage{epsfig}
\usepackage{graphicx}
\usepackage{color,ulem}

\makeatletter

\@addtoreset{equation}{section}
\makeatother
\newcommand{\eq}{\begin{equation}}
\newcommand{\eqx}{\end{equation}}
\newcommand{\eqs}{\begin{equation*}}
\newcommand{\eqsx}{\end{equation*}}
\newcommand{\eqn}{\begin{eqnarray}}
\newcommand{\eqnx}{\end{eqnarray}}
\newcommand{\bea}{\begin{eqnarray}}
\newcommand{\eea}{\end{eqnarray}}
\newcommand{\alg}{\begin{align}}
\newcommand{\algx}{\end{align}}

\def\({\left(}
\def\){\right)}
\def\<{\left<}
\def\>{\right>}

\def\[{\left[}
\def\]{\right]}

\newcommand{\tr}{{\rm tr}}

\begin{document}

\begin{titlepage}
\vskip1cm
\begin{flushright}
\end{flushright}
\vskip0.25cm
\centerline{
\bf \Large  
  Information and Coarse-Graining in Eternal Black Holes 
} 
\vskip 0.6cm \centerline{\large \textsc{
 Dongsu Bak}}
\vspace{0.6cm} 
\centerline{\sl   Physics Department,
University of Seoul, Seoul 02504 \rm KOREA}
 \vskip0.5cm

 \centerline{
\tt{ (dsbak@uos.ac.kr) }}
  \vspace{1.5cm}

\centerline{ABSTRACT} \vspace{0.75cm} \noindent
{Recently it was shown  in the context of  the AdS/CFT correspondence that the classical gravity description inevitably involves coarse-graining of some degrees.
  In this note we clarify 
the nature of this coarse-graining both in gravity and field theory sides. We show that the information carried by the bulk classical gravity
 fluctuation is transferred to the 
coarse-grained nonclassical degrees through tiny interactions, which makes the recovery of information almost impossible. Of course, the total system is evolving unitarily
fully preserving its information as dictated by the basic construct of the AdS/CFT correspondence.
}

\end{titlepage}

\section{Introduction}\label{sec1}
Based on the AdS/CFT correspondence \cite{Maldacena:1997re,Gubser:1998bc,Witten:1998qj}, one may  study bulk gravity dynamics in a rather precisely defined setup.  
Numerous aspects of the correspondence  have been checked for many examples in the zero temperature limit and  well understood by now. 
In this non thermal regime,
there are  no conceptual issues such as non-unitarity of description.   On the other hand,
 a precise understanding of the gravity description of thermal field theories is still lacking, which is relevant to the problem of the black hole 
information paradox \cite{Hawking:1976ra}.

In Ref.~\cite{Bak:2017xla}, rather general perturbations of BTZ black hole \cite{Banados:1992wn} are constructed 
which are dual to state 
deformations of the thermofield double theory \cite{Takahasi:1974zn}.
This construction may be viewed as 
a realization of  micro thermofield deformations of the BTZ geometry and may serve as an ideal setup to clarify  the issues of  black hole dynamics in AdS spacetime.
The perturbation makes one-point function of operator nonvanishing initially. This one-point function decays exponentially in general, which is
contradicting with the unitarity of the CFT defined on $R \times S^{d-1}$. 
Furthermore the future horizon area of this classical geometry grows  in time  in general.
Interpretation of the corresponding geometric  entropy is the key issue which we would like to address in this note.  We would like to clarify the nature of gravity description 
by investigating the degrees relevant to the growing of gravitational entropy. 

In short, the gravity description of thermal field theories 
naturally involves a coarse-graining of certain degrees. We identify the coarse-grained degrees as nonclassical degrees such as Hawking radiation 
quanta, which  cannot be  seen in the classical gravity description.
The interaction between 
the classical gravity degrees with these coarse-grained nonclassical degrees is tiny but nonvanishing and hence they work as a dark sector from the view point of 
the classical gravity description. In this note, we shall describe  the development of entanglement between the two and transfer of 
information from the classical  gravity  to the coarse-grained degrees.  
We shall also discuss how one can regain the unitarity  from the view points of both
 the gravity and the boundary field theory. 

In section \ref{sec2}, we review the general perturbations of BTZ background studied in Ref.~\cite{Bak:2017xla} and the corresponding AdS/CFT 
correspondence \cite{Maldacena:2001kr}. The basic puzzles of entropy and unitarity of the gravity description are described in detail. We clarify the degrees responsible for coarse-graining of bulk gravity description.  
In section \ref{sec3}, we present a field theory model to explain the entropy growth of  the perturbation as forming entanglement  between the classical gravity
 and the coarse-grained degrees. We also discuss how one can regain unitarity in this model. 
In section \ref{sec4}, we explain what happens from the bulk view point. This represents the quantum version of ER = EPR \cite{
Einstein:1935tc,Einstein:1935rr,
Maldacena:2013xja} where ER bridges are formed
between the classical and nonclassical degrees.  The latter degrees cannot be seen from the view point of the classical gravity description.

\section{Eternal black holes and perturbation of states}\label{sec2}
\begin{figure}[th!]
\centering  
\includegraphics[width=4.5cm,clip]{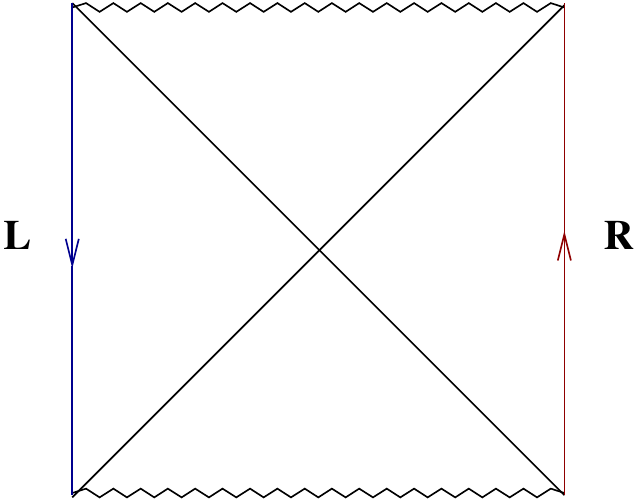} \hskip2cm
\includegraphics[width=4.5cm,clip]{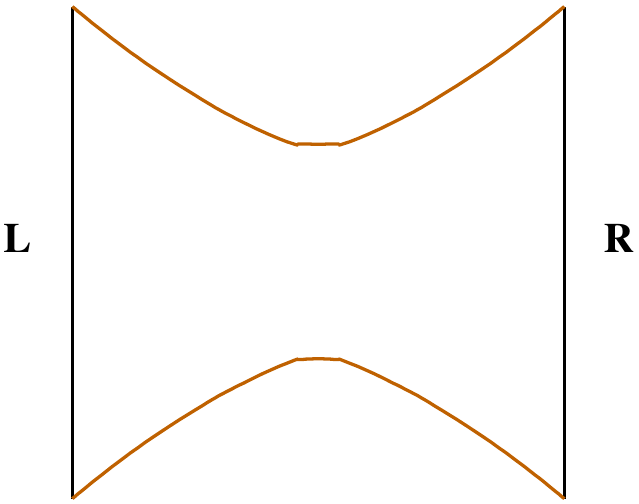}
\caption{\small On the left,  the Penrose diagram of the  BTZ black hole is depicted. Two copies of  a CFT live in the left and the right boundaries denoted by
$L$ and $R$ respectively.  On the right, we draw the ER bridge connecting the two boundaries. The left-right entanglement equals to the ER bridge of BTZ black hole connecting the left and the right boundaries.
} 
\label{figbtz}
\end{figure}

The eternal black hole in AdS spacetime is dual to the thermofield double of a CFT, which is maximally entangled but non interacting \cite{Maldacena:2001kr}.
The Hamiltonian of the total system is given by\footnote{The left and the right systems do not have to be the same in general, 
which leads to non maximal entanglement of the left and the right CFT's \cite{Bak:2007jm}. } 
\bea
H_{tfd}= H_L + H_R =H \otimes 1 + 1 \otimes H 
\eea
where there are absolutely no interactions between the left and the right CFT's\footnote{Interactions between the two CFT's can be introduced by a double trace deformation in \cite{Gao:2016bin}, which leads to a traversable wormhole solution.}. We denote the CFT Hamiltonian   by  
$H$. 
The initial unperturbed thermal vacuum state is given by a particularly prepared entangled state
\bea
|\Psi(0) \rangle =\frac{1}{\sqrt{Z}} \sum_{n,n'} \, \langle n | U | n' \rangle \, |n' \rangle \otimes | n  \rangle =\frac{1}{\sqrt{Z}} 
\sum_{n} \,  e^{-\frac{\beta}{2} E_n }\, |n\rangle \otimes | n \rangle
\label{initial}
\eea 
with a Euclidean evolution operator $U=U_0 = e^{-\frac{\beta}{2}  H}$ and $Z$ denoting the normalization constant. The entanglement here is maximal for a given temperature $T=1/\beta$. 
  In Figure \ref{figbtz}, we illustrate the Penrose diagram of the corresponding  BTZ black hole in 
three dimensions where two boundary spacetimes connected though an ER bridge representing the left-right entanglement. Here two copies of a 2d CFT live
on the left and the right boundary spacetimes  respectively.

This BTZ initial state can be deformed by introducing a mid-point insertion of operators as
\bea
U = e^{-\frac{\beta}{4}  H} e^{-\sum_k c_k O_k}  e^{-\frac{\beta}{4}  H} 
\label{initialnew}
\eea 
where $O_k$ denote general operators of the underlying CFT \cite{Bak:2017xla}. Then we evolve the system in time  
 in a standard manner
as
\bea
|\Psi(t) \rangle = e^{-i H_{tfd} \, t} \, |\Psi(0) \rangle 
\eea
The one-point function
\bea
\langle O_R \rangle =   \langle \Psi(t) | 1\otimes O |\Psi(t) \rangle
\eea
was computed  in \cite{Bak:2017xla}  to the leading order in its coefficient $c_k$ using the conformal perturbation theory and the two-point function \cite{KeskiVakkuri:1998nw, Maldacena:2001kr}
\bea
\frac{1}{Z} {\rm tr} O(t, \varphi) O(t', \varphi') e^{-\beta H }  
=
\sum^\infty_{m=-\infty}\frac{{\cal N}_\Delta}{
\left[\cosh \frac{2\pi}{\beta}( t - t')  - \cosh\frac{2\pi \ell}{\beta} (\varphi-\varphi'+2\pi m)- i \epsilon\right]^{\Delta}} 
\label{twopoint}
\eea
where $\Delta$ is the dimension of the scalar primary operator $O$ and ${\cal N}_\Delta$ \cite{Bak:2017rpp} is an appropriate normalization. Here for the sake of 
 an illustration, we
consider 2d CFT on $R \times S^1$ where we use coordinates $(t, \varphi)$ with $\varphi \sim \varphi+ 2\pi$. (Of course, our presentation can be generalized to 
other dimensions in a straightforward manner.) 
Then the field theory result agrees with the holographic computation of one-point function precisely \cite{Bak:2017xla}.  One finds that the resulting expression 
decays exponentially 
in time, which contradicts with the quantum Poincare recurrence theorem \cite{Dyson:2002pf}. Indeed the above expression of the two-point function is not exact but 
involves 
a large $N \, (\sim 1/\sqrt{G\ell^{1-d}})$ approximation\footnote{Here $\ell$ denotes the AdS radius and we shall consider the boundary $d$ dimensional CFT on $R \times S^{d-1}$ where the radius of the sphere is set to be $\ell$. For the simplicity of our presentation, $\ell$  is frequently set to be unity. 
One can be more precise about the large $N$  limit for the well known AdS$_5$/CFT$_4$ correspondence \cite{Maldacena:1997re} where the boundary CFT is given by ${\cal N}=4$ SU(N) super 
Yang-Mills theories for which $N$ is identified as the number of colors.}.  To understand this classical gravity approximation, we need to clarify the nature of the above large $N$ limit  in 
the dual field theories. 

Next we discuss the issue of entropy involved with the above deformations. 
Let us first
 introduce  the 
 reduced density matrix
\bea
\rho_R (t)= {\rm tr}_L \, |\Psi(t) \rangle  \langle \Psi(t) |
\eea
and then the corresponding  von Neumann entropy 
\bea
S_R = -\tr_{R\,} \rho_R (t) \log \rho_R (t) 
\label{vonr}
\eea
gives us  an entanglement measure
 of the left-right system. The perturbation in the above makes this left-right entanglement non-maximal initially.
One finds this L-R entanglement  is time independent  since one may undo the  similarity transformation of 
the time evolution of the reduced density matrix inside the trace. Since there are no interactions between the left right system, any net information of one side
cannot be transferred to the other side. This is consistent with the causal structure of the eternal black hole spacetime where 
the  left and the right boundaries are causally 
disconnected from each other. 

\begin{figure}[hb!]
\vskip0.5cm
\centering  
\includegraphics[width=4.5cm,clip]{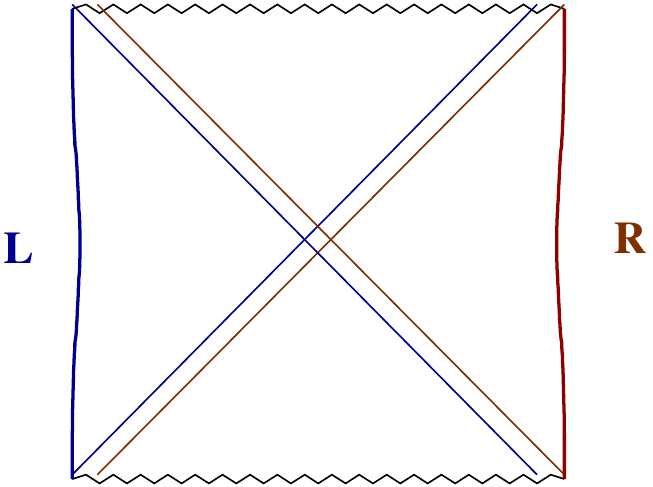}
\caption{\small  Penrose diagram of  perturbed  BTZ black hole is depicted here. The Hamiltonian of the system is not  perturbed but only initial state
is perturbed in a non maximally entangled manner. The diagram is elongated horizontally in such a way that the two sides are causally disconnected.
} 
\label{figtdbh}
\end{figure}

In Ref.\,\,\cite{Bak:2017xla}, it was also shown that,  for any perturbation of initial state in (\ref{initial}), 
there exists a  corresponding one-to-one  deformation of the eternal black hole spacetime.  See Figure \ref{figtdbh} for a Penrose diagram of  perturbed 
BTZ black hole.
These geometries show two general 
features. Any initial perturbations 
decay exponentially in time with  the time scale of  so called relaxation time scale 
$\beta$. In the gravity side, the perturbation outside horizon generically falls into the horizon and the corresponding horizon 
perturbations  decay away exponentially  in the time scale of the relaxation time scale.  This is generic in thermal AdS spacetime. With black holes, 
the decay is explained by the physics of quasi-normal modes which makes any perturbations of the black horizon decay away exponentially. This is consistent with
the field theory result using the large $N$ approximation in (\ref{twopoint}).

The second  is regarding entropy whose precise nature
is our primary concern in this note. In geometric side, the leading order gravitational entropy is given by the horizon area divided by $4 G$.  For instance we 
take the view point of the bulk observer outside of the future horizon from the right. The future horizon is nondecreasing monotonically in time 
reaching its asymptotic value after scrambling of the initial perturbation where the change of area $\Delta A$ is given by
\bea
 \Delta A = A(\infty)-A(0)
\eea   
which is finite 
and the function of the coefficients $c_k$.
In the corresponding gravitational entropy is given by 
\bea
S_{grav}(t)= \frac{A(t)}{4 G}   + S_\delta (t)
\label{grentropy}
\eea  
where $S_\delta$ denotes the contribution of order $G^0$ such as bulk entanglement entropy \cite{Bombelli:1986rw} which may be ignored in our discussion. 
In the late time, the area approaches
its final value $A(\infty)$  exponentially in time again with 
the relaxation time scale.  This leading term is the classical 
gravity contribution. 
It is rather clear that this entropy 
cannot be identified with the L-R entanglement entropy 
introduced in the above. The L-R entanglement entropy is time independent 
 whereas the above is time dependent.  

Understanding these two aspects will be our primary concern of this note. As was already noted in \cite{Bak:2007jm}, these two at least indicate that 
the classical gravity description cannot be fully fine-grained. If it were fully fine-grained, the exponential decay of perturbation 
is not possible 
since it is contradicting with the quantum Poincare recurrence theorem \cite{Dyson:2002pf}. It is also contradicting with the time independence of the L-R entanglement entropy 
which is dictated by the unitarity of the underlying system. Therefore it is rather clear that the classical gravity description cannot be fully fine-grained. This certainly 
implies that the classical gravity description is involving an inevitable coarse-graining. In this note, we would like to clarify the nature of gravity description in this respect. 
Of course the gravity description involves a large $N$ (or large central charge) approximation but the main question is which and how  degrees are 
coarse-grained in the gravity description. We shall not be fully general here and our main focus is on the perturbations around thermofield double state. 
The unperturbed case corresponds to the BTZ spacetime, which involves a maximal entanglement between the left and right system leading to an 
equilibrium  finite-temperature thermal system from the view point of one-sided 
observer.  The perturbations then make the system non-maximally entangled.

We divide each-sided system by $B$ and $C$ where $B$ is for the degrees responsible for the bulk classical gravity 
and $C$ the system of  the remaining degrees that  are coarse-grained by the classical gravity description. 
Let us choose  the right-side CFT  whose 
Hamiltonian is given by $H$ in the above. 
The part $C$ is the complement of $B$ with $R=B+C$ and forms an environment 
of $B$ in some sense. There has to be nonvanishing   interactions between $B$ and $C$. If there were no interactions, then the entropy would  be time 
independent  after integrating out those degrees of $C$. The interaction has to be as weak as $O(1/N)$. 
Otherwise it should be visible from the view point of
  the classical
 gravity. 

The part $C$ here is describing the quantum gravity degrees such as Hawking radiation of quanta whose existence is invisible from the view point of
 the classical gravity.
These include any of nonclassical degrees which are produced by 
$O(1/N)$ interactions.
Namely for instance one 
cannot see any classical  signal of such quanta from  the BTZ background though their mass may be included into the definition of the energy of the 
dual field theory.
Then $B$ is representing the part described by the classical gravity 
fluctuations. In this context, we shall refer the degrees of $C$ as nonclassical cloud (NCC) degrees\footnote{Here ``non-classicality" or ``quantum" refers 
to $1/N$ quantum gravity effects in the bulk that basically correspond to joining and splitting of strings  in the  string theory.},
which are coarse-grained from the view point of classical gravity 
description.   These degrees will be in thermal equilibrium with the black hole states forming a quantum cloud around the black hole since there is no net radiation 
for the case of BTZ black hole or the large AdS Schwarzschild black hole of other dimensions in general.
Below 
we shall present such a model of coarse-graining degrees and explain the above two characteristics of the eternal black hole dynamics. 

\section{Field theory model}\label{sec3}
In this section, we model the above phenomena observed in the gravity side. As we described already,  $B$ is representing 
the part described by the bulk fluctuation of the classical gravity 
that is one-to-one correspondence with the   
boundary deformation of the thermofield initial state with mid-point insertion of CFT
operators. These operators are basically dual to those bulk fields, which are basic elements of the gravity description.
We shall represent the corresponding field-theory  fluctuation above  the thermofield  vacuum by
\bea
|\psi_B \rangle =\sum_i \alpha_i |i \rangle 
\eea 
where $ |i \rangle $ is the eigenstate of   $H_B$ with eigenvalue   $ \epsilon_i$ with $i=1,2,\cdots$.  
The coarse-grained degrees in $C$ are responsible for the Poincare recurrence to happen in the full system $R=B+ C$. These degrees are excited in the black 
hole phase nonclassically  forming  the Hawking radiation cloud as we described previously. In the zero temperature limit, basically one may ignore these degrees
since their occupation numbers are practically zero. Hence the effect of $C$ disappears 
 in the zero temperature limit.  
It is not directly to do with 
higher derivative corrections since those higher derivative corrections are classical, which are well controlled in the AdS/CFT correspondence. 
Thus these higher derivative corrections 
will be included into the part $B$ of the classical gravity.  

Our assumption is rather mild here. First of all, the interaction between 
$B$ and $C$ has to be extremely weak. 
Hence the  degrees in $E=L+C$ should work as a dark sector from the view
point of the gravity system $B$ where the left side $L$ is completely dark. In the zero temperature limit of non thermal field theories, the occupation number of NCC degrees vanishes 
and their effect can be
 ignored completely. 
This is why the gravity description in the zero temperature limit is unitary and fully fine-grained.   
On the other hand, at finite temperature, 
the number of excited  NCC degrees in $C$  can be estimated as of order $N^0$ due to their semi-classical nature, which turns 
out to be still large enough\footnote{In the black hole phase, their number 
denoted by   $N_s$ 
 can be estimated as $N_s \sim f(\lambda, Tr)  {V_{S^{d-1}}}(T r)^{d-1}$ where $r$ is the radius of the sphere $S^{d-1}$,   ${V_{S^{d-1}}}$ the volume of a unit 
 $d-1$ sphere and $\lambda$ denotes some moduli 
 parameter such as  't Hooft coupling in 4d ${\cal N}=4$ SYM theories.  $f(\lambda, Tr)$ is counting the bulk fields whose Hawking radiation quanta are significantly excited.
Since  $rT \gg 1$ and $f(\lambda, Tr)$ is large  in the black hole phase, the number of excited  $NCC$ degrees has to be large.}.
Thus many such NCC degrees are excited which we label by $I=1, 2, \cdots, N_s $.  
 The  number of relevant states for the full environment  of $E=L+C$ is of order $M^{N_s} \sim e^{\alpha N^2}$ where
we assume  
there are $M$ states for each NCC degree. 

Then, for the $I$-th 
NCC degree, the relevant state is given by
\bea
|\psi \rangle_I = \sum^M_{m=1} c^I_m \,  | m \rangle_I 
\eea
where we use the basis defined by  $H_I  | m \rangle_I =\frac{1}{M} e_m | m \rangle_I$ where $M$ is as large as $M \sim e^{\alpha N^2/N_s}$.
The full relevant Hilbert space  will be described by the tensor product
\bea
| \psi_E \rangle  = \prod^{N_s}_{I=1} |\psi \rangle_I 
\eea
We take the initial state of $E$ as
\bea
|\psi \rangle_I (0)=\frac{1}{\sqrt{M}} \sum^M_{m=1} e^{i \varphi^I_m} \,  | m \rangle_I 
\eea  
with random phases  $\varphi^I_m$ where all eigenstates are equally probable.
The interaction Hamiltonian is given by \footnote{We assume here the interaction Hamiltonian is diagonalized by the basis $|i \rangle \otimes |m\rangle_I$. This assumption is not necessary and  just for the simplicity of our presentation.}
\bea
H_{int} =  \frac{g_0}{N'} H_B \otimes \sum_I H_I
\eea
where we assume that  $N'$ goes to infinity as  $N_s$ tends to infinity.
We shall assume that $H_B$ responsible of interactions is non-degenerate. 
Thus we begin with  an initial state 
\bea
\sum_i \alpha_i\, |i\rangle \otimes |\psi_E \rangle (0)
\eea
The entanglement entropy between $B$ and $E$ is zero since the reduced density matrix tracing over $E$ is 
still pure. Then the time evolution of the system is given by
\bea
\sum_i \alpha_i \,|i\rangle \otimes |\psi^i_E \rangle (t)
\eea
where one has
\bea
| \psi^i_E \rangle (t)  = \prod^N_{I=1} |\psi^i \rangle_I (t) 
\eea
with
\bea
|\psi^i \rangle_I (t)=\frac{1}{\sqrt{M}} \sum^M_{m=1} e^{i \varphi^I_m-i \frac{g_0 \epsilon_i e_m}{N' M}t} \,  | m  \rangle_I   
\label{sistate}
\eea
Then the overlap can be evaluated as
\bea
\langle \psi^i  |\psi^{j} \rangle_I  =\frac{1}{M} \sum^M_{m=1} e^{i g_0 \frac{ \epsilon_i -\epsilon_j}{N'} \frac{e_m}{M}t} 
\eea
Since $M$ is large enough, we can turn this  to an integral 
\bea
\langle \psi^i  |\psi^{j} \rangle_I  =\int^{\infty}_{-\infty} dx \, D(x) \, e^{i  g_0 \frac{ \epsilon_i -\epsilon_{j}}{N'}x t }
\eea
where we introduce the normalized density of state $D(x)$ by
\bea
D(x)= \frac{1}{M} \sum^M_{m=1} \delta \Big(x-\frac{e_m}{M}\Big)
\eea
which satisfies the normalization
\bea
\int^{\infty}_{-\infty} dx\,  D(x) =1
\eea
For instance choosing $D (x)=D_0(x)=\frac{1}{2 \pi} \Theta(\pi-x)\Theta(x+\pi)$ leads to
\bea
\langle \psi^i  |\psi^{j} \rangle_I  = \frac{\sin \pi  \frac{ \epsilon_i -\epsilon_{j}}{N'}g_0 t }{\pi  \frac{ \epsilon_i -\epsilon_{j}}{N'} g_0t}
\eea
This leads to the overlap 
\bea
\langle \psi_E^i  |\psi_E^{j} \rangle = e^{-\frac{\pi^2}{6} (\epsilon_i -\epsilon_{j})^2 g_0^2 t^2 }
\eea
where we assume $ N_s={N'}^2$  and take the large $N_s$ limit. 
This  decay rate is too fast since in the gravity side we have just exponential decay rate in the late time. 
Hence this example is not consistent with our bulk gravity dynamics  described in section \ref{sec2}.

 In this example, 
 let us take
$M$ and $N'$ to be finite and further assume $M$ is even integer. By taking $e_m =m -M/2$ and $\epsilon_i = i$, one finds $D(x)=D_0(x)$ in the large $M$ limit.
With this choice of $e_m$ and $\epsilon_i$, the state in  (\ref{sistate}) will return to its initial state as
\bea
 |\psi^i \rangle_I (0)  =|\psi^i \rangle_I (t_{rec})
\eea
where $t^E_{rec}= \pi N'M/g_0$ 
with $t^E_{rec}$ denoting
the Poincare 
recurrence time of the system. If one allows a slight random variation of   multiplicity $M$ over $I$ with its mean 
value $\langle M \rangle =e^{\alpha N^2/N_s}$ and assumes $N'=N_s$, 
the recurrence time scale 
becomes $t^E_{rec} \sim e^{N_s\log N_s} e^{ \alpha  N^2 } \sim  e^{S_E} $. 

\begin{figure}[th!]
\centering  
\includegraphics[height=3.5cm]{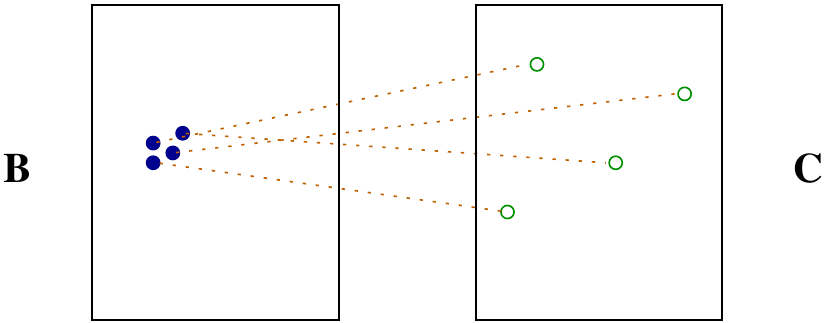}
\caption{\small  $B$ and $C$  represent respectively the classical gravity degrees and the remaining  nonclassical degrees in $R$. EPR pairs between 
$B$ and $C$ are formed by entanglement, which are represented by the dotted lines. The interaction is small and the location of degrees in $C$ are random and $C$
is embedded in $E$. 
} 
\label{figbcfield}
\end{figure}

To consider a more realistic case, we choose 
\bea
D(x) =\frac{1}{\pi}\frac{a_0}{x^2+ a_0^2}
\eea
together with $N'=N_s$.
Then one has 
\bea
\langle \psi^i  |\psi^{j} \rangle_I  =e^{-| \frac{ \epsilon_i -\epsilon_{j}}{N'}a_0 g_0t|}
\eea
This leads to the overlap 
\bea
\langle \psi_B^i  |\psi_B^{j} \rangle = e^{-|(\epsilon_i -\epsilon_{j})a_0g_0 t | }
\eea
This is describing a typical decoherence \cite{Zurek:2003zz}
and consistent with the late time behavior of the perturbations in \cite{Bak:2017xla}. We take $a_0$ to be an order of typical thermal energy scale as $a_0 = b_0 T$.
For example consider the case 
\bea
|\psi_B \rangle = \frac{1}{\sqrt{2}}\big( |0 \rangle +  |1 \rangle  \big)
\eea
Then the reduced density matrix is obtained as
\bea 
\rho_B =\frac{1}{2} \Big[ |0 \rangle \langle 0 | + |1 \rangle \langle 1 | + \chi (t)  (  |0 \rangle \langle 1 |+  |1 \rangle \langle 0 | )  \Big]
\eea
where $\chi(t) = e^{-k_0 \frac{t}{\beta}}$ with $k_0 =|( \epsilon_1 -\epsilon_{0})b_0 g_0| $ for $t >0$.
The corresponding entanglement entropy is given by
\bea
\
\Delta S_B(t) = \log 2 -\frac{1}{2} (1+ \chi(t)) \log (1+ \chi(t)) -\frac{1}{2} (1- \chi(t)) \log (1- \chi(t))
\eea
with $\Delta S_B(t) =S_B(t)-S_B(0)$. (See below for the definition of $S_B$ in the full thermofield double system.)
One finds initially there is no entanglement as $\Delta S_B(0)=0$. Then the information is  leaking out from the system $B$ to the dark cloud environment $C$
though the interaction between $B$ and $C$ is negligibly small. The final entanglement entropy approaches the maximal value $\log 2$ exponentially
as a time scale of typical relaxation time scale whose precise value depends on the properties of system $B$.  This feature is true for any $|\psi_B\rangle (0)$. 
At the initial moment of perturbation,  there is no entanglement between $B$ and $C$. Then through the interactions which is extremely small, the information 
leaks out from $B$ to $C$ by forming EPR pairs as described in Figure \ref{figbcfield}.

Since the system $E=B+C$ is huge, it is practically impossible to collect this entanglement back to the original 
non entangled  state by performing  simple operations.  Thus in the gravity description, there is an effective loss of information through the entanglement 
between $B$ and $C$. This is the story well before the Poincare recurrence time scale.

The typical recurrence time scale becomes $t^E_{rec}\sim e^{ S_E}\sim e^{S_{grav}}$. 
It is clear that the initial information of system $B$ can be recovered if one waits 
for the order of the recurrence time scale. In this sense there is certainly no loss of information by the entanglement transfer since 
the transferred information will be regained if one waits for  enough time  that is order of the recurrence time scale.

\section{Bulk interpretation} \label{sec4}
In this section we shall describe how the above transfer of information from $B$ to $C$ looks like from the view point of the bulk
side. 
The expansion in the gravity side is organized as follows. There are saddle-point solutions each of which  
is weighted 
by the probability $\frac{1}{Z} e^{-I}$ where $I$ is the on-shell action and $Z$ is the full partition function\footnote{Of course each saddle point receives 
quantum gravity corrections of $1/N$ expansions}. 
Hence the entropy for instance is given by
\bea
S_{grav}=\langle \hat{S}_{grav} \rangle = \frac{1}{Z}\sum_\alpha e^{-I_\alpha} S^{grav}_\alpha
\eea
 where $\alpha$ is labeling each saddle-point and $S^{grav}_\alpha$ is the entropy in (\ref{grentropy}) associated with the saddle-point.
This entropy is well defined at least for the static case in which the system is in thermal equilibrium. As we showed already, the leading contribution of the 
above entropy is from that of the deformation of BTZ spacetime if the temperature is larger than the Hawking-Page transition temperature 
$T_{HP}= \frac{1}{2\pi \ell}$ \cite{Hawking:1982dh}. Then the area and the entropy grows in time, which implies that the  corresponding entropy in field theory 
side   cannot be identified with the L-R entanglement entropy $S_R$ in (\ref{vonr}) obtained by tracing over the CFT on the left side. As
we said before, this R-L entanglement is time-independent  since there is no interaction between the left and the right sides. This then should be identified with 
the entanglement entropy where we further trace over the degrees in $C$ which is invisible from the view point of gravity description. We introduce
further reduced density matrix by
\bea
\rho_B = \tr_C \, \rho_R
\eea
Hence  the entanglement entropy between  $B$ and $E=C+L$ (with $R= B+C$)
\bea
S_{B}= -\tr_B \rho_B \log \rho_B
\eea
can be identified with the gravitational entropy $S_{grav}$ in (\ref{grentropy}). 
But there is a nonvanishing interaction between $B$ and $C$ as emphasized previously. Thus there is a leak of information from $B$
to $C$, which leads to the increase of the entropy $S_{grav}$. This leak happens in the form of developing entanglement between 
$B$ and $C$ as described in the previous section.  
\begin{figure}[th!]
\centering  
\includegraphics[height=4cm]{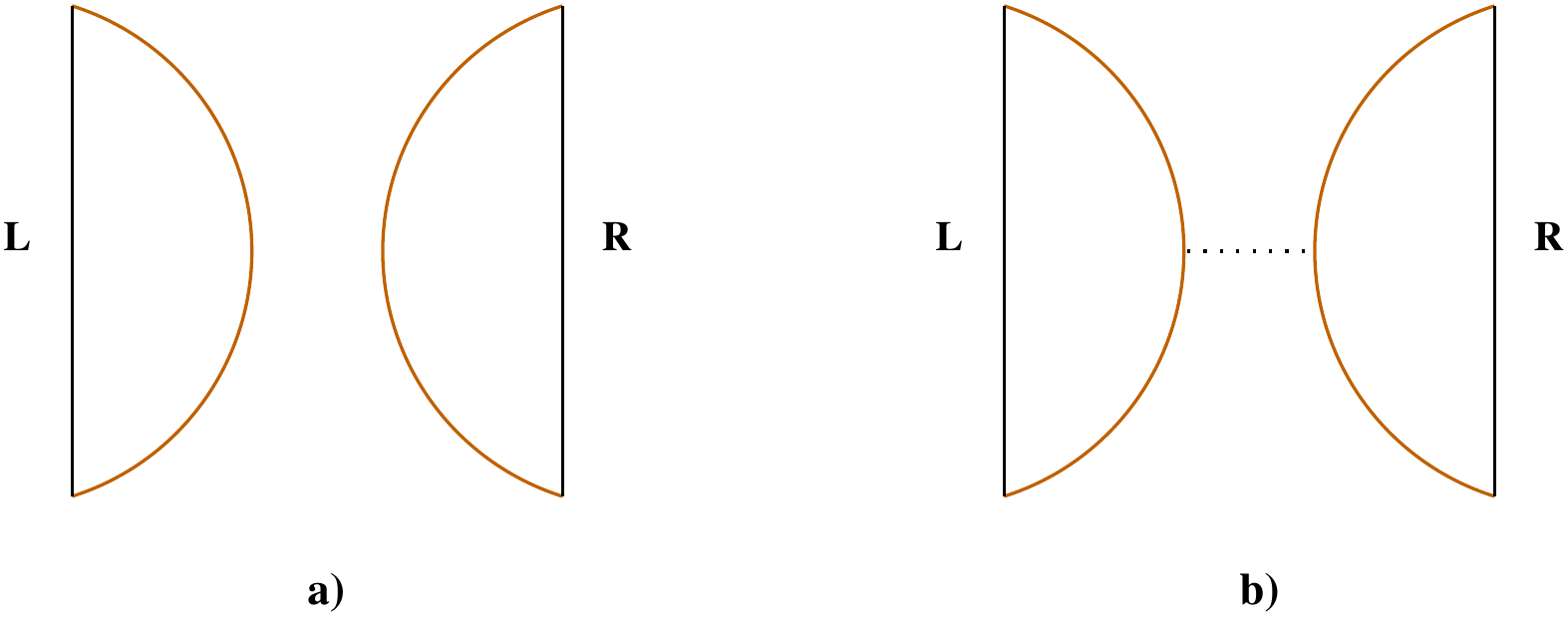}
\caption{\small In the thermal AdS phase the left and right sides are classically disconnected. But there is a contribution from graviton exchange between them 
semiclassically which is responsible for the entanglement between the left and the right sides. This is depicted in b) where the dotted line represents the graviton 
exchange.
} 
\label{figthads}
\end{figure}
Hence according to the ER = EPR conjecture \cite{Maldacena:2013xja}, 
the corresponding ER bridge connecting $B$ to $C$ should develop in the gravity description. However
our solutions found in \cite{Bak:2017xla}  do not show any signature of the development of ER bridge.  This ER bridge ought to be nonclassical and
cannot be seen in the classical gravity description.
A similar behavior may be found in other examples.  For instance consider the thermal AdS geometry that gives a dominant contribution to the partition function or the sum over geometries for $T < T_{HP}$, 
which is basically  the Hawking-Page transition \cite{Hawking:1982dh}.
There are two thermal AdS geometries for the left and the right  CFT's separately. More precisely, as described in 
\cite{Maldacena:2001kr}, two Lorentzian thermal AdS spaces are  connected by a half of Euclidean thermal AdS solution in order to provide the 
initial state in  (\ref{initial}).  Thus this geometry is still dual to the thermofield double dynamics. It is clear that there is a nonvanishing L-R entanglement 
that is given by (\ref{vonr}) once the temperature is nonzero. The corresponding entanglement entropy is of higher order in $G$ 
and hence cannot be seen in the classical 
 gravity
description. 
Indeed the two Lorentzian geometries of the left and the  right sides are completely disconnected as depicted in Figure \ref{figthads}. Hence at the level of solution, the ER
bridge does not appear, which is of course not a contradiction. There are many other examples of this type:  The connected string solution in Figure\,\,1c of \cite{Bak:2007fk}
appears as two completely separate strings at the level of classical description.

 \begin{figure}[th!]
\centering  
\includegraphics[height=3.7cm]{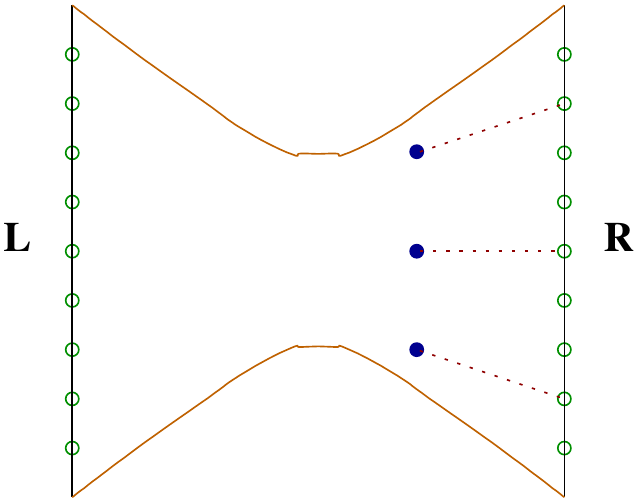}
\caption{\small On the boundaries, the green dots represent the degrees of $C$ not covered by the classical gravity system. The rest represent the degrees in $B$ of  the classical gravity  including blue 
dots. The left-right entanglement is reduced to a non maximal one by the perturbations. 
The bulk degrees in $B$ (denoted by blue dots) is connected by the ER connecting $C$ (denoted by green dots) which is nonclassical. This effect is of order $1/N$ and cannot be seen in the 
classical gravity solutions.
} 
\label{figbc}
\end{figure}

 Similarly in our example, the ER bridge is not seen at the level of classical description.
 The relevant degrees in $B$  fall gradually into the horizon that is responsible for the growth of the horizon area. Thus the entanglement is between 
 these behind-horizon degrees and the corresponding degrees in $C$. The degrees in $C$ is not accessible from the classical gravity description. 
 The connection is the quantum gravity fluctuation 
 which is the semiclassical picture of the relevant ER bridge.   
   We illustrate this in Figure \ref{figbc}.

Finally let us comment on the issue of the recurrence in the bulk.
In the classical gravity description, the large $N$ limit 
is already taken, so there is basically no way to see this even in principle. But through the higher order effect in $1/N$ such as Hawking radiation, 
regaining of the transferred entanglement is 
allowed in principle, which can be a possible resolution of the black hole  information paradox. 
But here we are dealing with an eternal black hole. Unlike the case of evaporating black hole in the flat space, 
the regaining time scale will be as large as $t_{rec}\sim  e^{\alpha N^2}\sim e^{S_{grav}}$.
At the final stage of recurrence, the entanglement between $B$ and $C$ is reduced and the system is back to the original state.
In the time dependent perturbation of  black hole in \cite{Bak:2017xla}, this return actually happens when $t < 0$, in which the entanglement between $B$ and $C$ is decreasing through 
interactions. Not that there is a failure of description at $t= \pm \infty$, which appears as the orbifold singularity in the three dimensional  BTZ 
black hole. We view this as a simple failure of description, so in the bulk the recurrence will occur if one waits for long enough time.

\section*{Acknowledgement}
This work was
supported in part by
NRF Grant 2017R1A2B4003095.



\begin{thebibliography}{99}\label{bib}


\bibitem{Maldacena:1997re} 
  J.~M.~Maldacena,
  ``The Large N limit of superconformal field theories and supergravity,''
  Int.\ J.\ Theor.\ Phys.\  {\bf 38}, 1113 (1999)
  [Adv.\ Theor.\ Math.\ Phys.\  {\bf 2}, 231 (1998)]
  [hep-th/9711200].



\bibitem{Gubser:1998bc} 
  S.~S.~Gubser, I.~R.~Klebanov and A.~M.~Polyakov,
  ``Gauge theory correlators from noncritical string theory,''
  Phys.\ Lett.\ B {\bf 428}, 105 (1998)
  [hep-th/9802109].



\bibitem{Witten:1998qj} 
  E.~Witten,
  ``Anti-de Sitter space and holography,''
  Adv.\ Theor.\ Math.\ Phys.\  {\bf 2}, 253 (1998)
  [hep-th/9802150].

\bibitem{Hawking:1976ra} 
  S.~W.~Hawking,
  ``Breakdown of Predictability in Gravitational Collapse,''
  Phys.\ Rev.\ D {\bf 14}, 2460 (1976).
  doi:10.1103/PhysRevD.14.2460


\bibitem{Bak:2017xla} 
  D.~Bak, C.~Kim, K.~K.~Kim and J.~P.~Song,
  ``Holographic Micro Thermofield Geometries of BTZ Black Holes,''
  JHEP {\bf 1706}, 079 (2017)
  [arXiv:1704.01030 [hep-th]].



\bibitem{Banados:1992wn} 
  M.~Banados, C.~Teitelboim and J.~Zanelli,
  ``The Black hole in three-dimensional space-time,''
  Phys.\ Rev.\ Lett.\  {\bf 69}, 1849 (1992)
  [hep-th/9204099].



\bibitem{Takahasi:1974zn} 
  Y.~Takahasi and H.~Umezawa,
  ``Thermo field dynamics,''
  Collect.\ Phenom.\  {\bf 2}, 55 (1975).




\bibitem{Maldacena:2001kr} 
  J.~M.~Maldacena,
  ``Eternal black holes in anti-de Sitter,''
  JHEP {\bf 0304}, 021 (2003)
  [hep-th/0106112].

\bibitem{Einstein:1935tc} 
  A.~Einstein and N.~Rosen,
  ``The Particle Problem in the General Theory of Relativity,''
  Phys.\ Rev.\  {\bf 48}, 73 (1935).

\bibitem{Einstein:1935rr} 
  A.~Einstein, B.~Podolsky and N.~Rosen,
  ``Can quantum mechanical description of physical reality be considered complete?,''
  Phys.\ Rev.\  {\bf 47}, 777 (1935).



\bibitem{Maldacena:2013xja} 
  J.~Maldacena and L.~Susskind,
  ``Cool horizons for entangled black holes,''
  Fortsch.\ Phys.\  {\bf 61}, 781 (2013)
  [arXiv:1306.0533 [hep-th]].


\bibitem{Bak:2007jm} 
  D.~Bak, M.~Gutperle and S.~Hirano,
  ``Three dimensional Janus and time-dependent black holes,''
  JHEP {\bf 0702}, 068 (2007)
  [hep-th/0701108];
  D.~Bak, M.~Gutperle and A.~Karch,
  ``Time dependent black holes and thermal equilibration,''
  JHEP {\bf 0712}, 034 (2007)
  [arXiv:0708.3691 [hep-th]].



\bibitem{Gao:2016bin} 
  P.~Gao, D.~L.~Jafferis and A.~Wall,
  ``Traversable Wormholes via a Double Trace Deformation,''
  arXiv:1608.05687 [hep-th].


\bibitem{KeskiVakkuri:1998nw} 
  E.~Keski-Vakkuri,
  ``Bulk and boundary dynamics in BTZ black holes,''
  Phys.\ Rev.\ D {\bf 59}, 104001 (1999)
  [hep-th/9808037].




\bibitem{Bak:2017rpp} 
  D.~Bak and A.~Trivella,
  ``Quantum Information Metric on R X S(d-1),''
  JHEP {\bf 1709}, 086 (2017)
  [arXiv:1707.05366 [hep-th]].

\bibitem{Dyson:2002pf} 
  L.~Dyson, M.~Kleban and L.~Susskind,
  ``Disturbing implications of a cosmological constant,''
  JHEP {\bf 0210}, 011 (2002)
  [hep-th/0208013].




\bibitem{Bombelli:1986rw} 
  L.~Bombelli, R.~K.~Koul, J.~Lee and R.~D.~Sorkin,
  ``A Quantum Source of Entropy for Black Holes,''
  Phys.\ Rev.\ D {\bf 34}, 373 (1986).




\bibitem{Zurek:2003zz} 
  W.~H.~Zurek,
  ``Decoherence, einselection, and the quantum origins of the classical,''
  Rev.\ Mod.\ Phys.\  {\bf 75}, 715 (2003).


\bibitem{Hawking:1982dh} 
  S.~W.~Hawking and D.~N.~Page,
  ``Thermodynamics of Black Holes in anti-De Sitter Space,''
  Commun.\ Math.\ Phys.\  {\bf 87}, 577 (1983).

\bibitem{Bak:2007fk} 
  D.~Bak, A.~Karch and L.~G.~Yaffe,
  ``Debye screening in strongly coupled N=4 supersymmetric Yang-Mills plasma,''
  JHEP {\bf 0708}, 049 (2007)
  [arXiv:0705.0994 [hep-th]].


\end{thebibliography}
\end{document}